\documentstyle[aps,aipbook,floats]{revtex}
\begin{document}

\input psfig.sty

\rightline{OCIP/C-95-5}
\rightline{hep-ph/9505252}
\rightline{April  1995}

\title{Quartic Gauge Boson Couplings}

\author{Stephen Godfrey\thanks{e-mail: godfrey@physics.carleton.ca}}
\address{Department of Physics, Carleton University, \\
Ottawa, CANADA K1S 5B6}

\maketitle

\begin{abstract}
Quartic vertices provide a window into
one of the most important problems in particle physics;
the understanding of electroweak symmetry breaking.
I survey the various processes that have been proposed to study
quartic gauge boson couplings at future
$e^+e^-$, $e\gamma$, $\gamma\gamma$, $e^-e^-$, and $pp$ colliders.
For the lowest dimension operators that do not include photons,
it appears that the LHC will
provide the most constraining measurements.  However,
precision measurements at high energy $e^+e^-$
colliders involving  $W^+W^-$ rescattering are also quite sensitive
to the effects of a strongly interacting weak interaction.
For quartic couplings involving
photons, $\gamma\gamma$ collisions appear to be the best place to
measure these couplings.
Measurements using gauge boson production in $e\gamma$ collisions are
almost as precise as the $\gamma\gamma$  processes with
$e^+e^-\to VVV$ about an order or magnitude less sensitive.

\end{abstract}

\section{Introduction}

The non-Abelian gauge nature of the standard model predicts, in
addition to the trilinear $WWZ$ and $WW\gamma$ couplings (TGV's), quartic
gauge boson couplings (QC's).
The strength of the couplings is set by the universal gauge
couplings of the $SU(2)$ local gauge symmetry.
In the standard model there are only three
quartic couplings which necessarily involve at least two charged
$W$'s;  $W^+ W^- W^+ W^-$, $W^+ W^- Z Z$, and $W^+ W^- \gamma
\gamma$.  Although
the $ZZZZ$ vertex is not present in the SM it is present at tree
level via Higgs
exchange while the $\gamma\gamma ZZ$ vertex is only produced at loop
level in the Standard Model.

The trilinear and quartic couplings probe different aspects of the
weak interactions.  The trilinear couplings
test the non-Abelian gauge structure where deviations from the SM can
result from integrating out heavy particles in loops \cite{aihara}.
In contrast, the quartic couplings can be regarded as
a window on electroweak symmetry breaking.
Recall that the
longitudinal components of the $W$ and $Z$ are Goldstone bosons.
The quartic couplings of gauge bosons therefore represent a connection
to the scalar sector of the theory.  The QC's would arise
as a contact interaction manifestation of heavy particle exchange.

It is quite possible that the quartic couplings deviate from their
SM values while the TGV's do not.  For example, the BESS model is a
non-linear realization of symmetry breaking where new structures not
present at tree level appear in 4$W$ couplings\cite{bess}.
There are models with a heavy scalar singlet interacting with the
Higgs sector which do not affect the $\rho$ parameter nor the TGV's
but do change the $4W$ vertex \cite{hill}.

Thus, if the mechanism for electroweak symmetry breaking does not
reveal itself through the discovery of new particles such as
the Higgs boson, supersymmetric particles, or technipions it is quite
possible that anomalous quartic couplings could be our first probes
into this sector of the electroweak theory.

While considerable effort has been expended to study the
trilinear couplings, the quartic couplings are only starting to
receive much attention. In this
contribution I review recent developments in the study of quartic
couplings and attempt to summarize the current status of this
subject.  In the next section I describe the
effective Lagrangians relevant to QC's. I will then describe
various processes  that have been proposed to study
quartic couplings using a wide variety of colliding
particles: $pp$, $e^+e^-$, $e\gamma$, $\gamma\gamma$, and $e^-e^-$.
In the final section I summarize these results and also add
some comments as to where the subject can benefit from further work.

\section{Effective Lagrangians and Quartic Couplings}

The formalism of effective Lagrangians provides a well-defined
framework for investigating the physics of anomalous couplings and
electroweak symmetry breaking \cite{boudjema,bagger93,baillargeon}.
The infinite set of terms in
${\cal L}_{eff}$ can be organized in an energy expansion where at low
energy only a
finite number of terms will contribute to a given process.  At
higher energies more and more terms become important until the
whole process breaks down at the scale of new physics. One focuses
on the leading operators in the expansion.

Quartic operators can either be associated with trilinear couplings
or can be genuinely quartic.  The former type is described by:
\begin{equation}
{\cal L}^{WW\gamma} = -i e {\lambda_\gamma \over M_W^2} F^{\mu\nu}
W_{\mu\alpha}^\dagger W_\nu^\alpha
\end{equation}
This operator generates
$WW\gamma\gamma$ couplings with strength
$e^2\lambda_\gamma$ in addition to  $WW\gamma$ couplings.
These vertices are not likely to be
particularly interesting as the parameter $\lambda_\gamma$ will
already be constrained from other processes such as $e^+e^-\to WW$
where the TGV contributes but the QC does not appear \cite{aihara}

We will restrict our discussion to the more interesting, genuinely
quartic couplings.  We
concentrate on the lowest dimension operators that can contribute to
a given vertex.  We impose custodial $SU(2)$, which is satisfied
to high precision by the nearness of the $\rho$ parameter to unity,
and $U(1)_{em}$ for the operators involving photons.
There are 2 (equivalent) parametrizations which
have appeared in the literature.  We will begin by describing these
parametrizations.

\subsection{General Parametrization}

This parametrization was introduced by B\'elanger and Boudjema
\cite{belanger92a,belanger92b}.
There are only two dimension four operators.
They do not involve photons since $U(1)_{em}$ requires derivatives
which would result in a higher dimension operator.  Imposing
$SU(2)_C$ the two dimension four operators are given by:
\begin{eqnarray*}
{\cal L}^o_4 & = & {1\over 4} g_o g_W^2 (\vec{W}_\mu
	\cdot\vec{W}^\mu)^2 \\
	& & \to g_o g_W^2
		[(W_\mu^+ W^{-\mu})(W_\nu^+ W^{-\nu})
	+ {1\over c_w^2} W_\mu^+ W^-_\mu Z^\nu Z_\nu
		+{1\over{4c_w^2}} {Z^\mu Z_\mu Z_\nu Z^\nu} ]  \\
	{\cal L}^c_4 & = & {1\over 4} g_c g_W^2
		(\vec{W}_\mu \cdot\vec{W}^\nu) (\vec{W}^\mu\cdot\vec{W}_\nu)\\
	& & \to  g_c g_W^2
[{1\over 2}(W_\mu^+ W^{-\mu} W_\nu^+ W^{-\nu}+W_\mu^+ W^{+\mu} W_\nu^-
W^{-\nu}) \\
	& & \qquad\qquad +{1\over c_w^2} W_\mu^+ W^{-\nu} Z^\mu Z_\nu
		+{1\over{4c_w^2}} Z^\mu Z_\mu Z_\nu Z^\nu ]
\end{eqnarray*}
These operators involve the maximum number of longitudinal modes.  These
are the most important manifestations of an alternative symmetry
breaking scenario.
Note that the $ZZZZ$ vertex does not appear in the SM and
$W^+W^-W^+W^-$ cannot be probed via 3 boson production in $e^+e^-$.
Photons do not appear in these genuine QC's.

The first operator can be thought of as parametrizing heavy neutral
scalar exchange so that we can make the connection:
\begin{equation}
g_o \propto \kappa^2 \left({M_W^2\over \Lambda^2}\right)
\end{equation}
where $\kappa$ is the strength of coupling in the $W$ system.  Heavy
Higgs exchange, at tree level in the SM, gives $\kappa$ of order 1.
In this case $g_o \simeq 0.2$ corresponds to $M_H=\Lambda \sim 180$~GeV which
would most likely be observed directly at a high energy collider
invalidating this approach.  On
the other hand, taking a Higgs mass of 1~TeV yields a contact term of
strength $g_0\simeq 6\times 10^{-3}$.  Thus, to see the effect of a
heavy scalar as a deviation to the QC requires very precise
measurements.

For the case of scalar exchange $g_o>0$.
In the second operator the $WWZZ$ vertex corresponds to heavy charged
scalar exchange so that we
could associate it with a triplet of heavy scalars.
A specific case of interest is when $g_o=-g_c = g_s<0$
which could parametrize heavy vector particle exchange which might
arise in theories like technicolour.  In this case the $4Z$
couplings cancel and the net effect is a rescaling of the SM $4W$
vertex.  Bounds on $g_s$ therefore determine the precision with which $4W$
couplings could be measured.

To introduce photons we have to go to dimension 6 operators.  We
only consider these dim-6 operators since they result
in the largest phase space and are therefore most likely to give the
largest deviations.
%They contribute to three-boson
%production processes in $e^+e^-$ collisions
%not described by the dimension-4 operators and
%the real $\gamma\gamma$ collisions of the  laser mode of the NLC:
%\begin{eqnarray}
%e^+e^- & \to &  W^+ W^- \gamma , \;  ZZ \gamma , \; Z\gamma\gamma \nonumber\\
%\gamma\gamma & \to & W^+ W^- , \;  ZZ
%\end{eqnarray}
Imposing $SU(2)_C$ and $U(1)_{QED}$
and restricting the phenomenological analysis to the $C$ and $P$
conserving operators with a maximum of two photons
involves the $\gamma\gamma W^+ W^-$ and $\gamma\gamma Z Z$ vertices
described by the operators:
\begin{eqnarray}
{\cal L}_6^0 & = & -{\pi \alpha \over {4\Lambda^2}} a_0 F_{\alpha\beta}
F^{\alpha\beta} (\vec{W}_\mu \cdot \vec{W}^\mu ) \\
{\cal L}_6^c & = & -{\pi \alpha \over {4\Lambda^2}} a_c F_{\alpha\mu}
F^{\alpha\nu} (\vec{W}_\mu \cdot \vec{W}^\nu )
\end{eqnarray}
Where $\vec{W}_{\mu}$ is an $SU(2)$ triplet and $F^{\mu\nu}$ and
$\vec{W}^{\mu\nu}$ are the $U(1)_{em}$ and $SU(2)$ field strengths
respectively.  Both operators
have contributions from loops but the first can originate from
heavy neutral scalar exchange while the second can arise from
charged scalars.
Note that the $SU(2)$ gauge symmetry predicts
that $\gamma\gamma ZZ$ does not appear in the SM.
The custodial symmetry imposed on these couplings means that, in
leading order in $s$, they contribute in the same way to the
$\gamma\gamma \to WW$ and to $\gamma\gamma\to ZZ$.

There is an additional operator which gives
a $W^+W^-Z\gamma$ vertex \cite{eboli94}:
\begin{eqnarray}
{\cal L}^n & = & {i\pi \alpha \over {4\Lambda^2}} a_n \varepsilon_{ijk}
W_{\mu\alpha}^i W_\nu^j W^{k\alpha}  F^{\mu\nu}
\end{eqnarray}

The parameter $\Lambda$ is an unknown ``new physics''
scale which is often taken to be $M_W$. This is a little misleading as
$\Lambda$ represents the scale of new physics which one might expect
to be ${\cal O}(1)$~TeV.  One should keep this in mind when gauging
the sensitivity of various experiments to the parameters $a_i$.  To
facilitate the comparison of different processes I have taken
$\Lambda=1$~TeV, rescaling results where necessary.

${\cal L}_0$ and ${\cal L}_c$ affect the value of $\Delta r$ and
therefore contribute to the $S$ and $T$ parameters \cite{peskin}
leading to the rather weak one-sigma constraints\cite{eboli94}:
\begin{eqnarray}
%-4.5 & < a_0 < & 0.64 \nonumber \\
%-11 & < a_c < & 5.8.
-700 & < a_0 < & 100 \nonumber \\
-1700 & < a_c < & 900.
\end{eqnarray}
There are no similar low energy constraints on $a_n$.

\subsection{Non-Linear Realization}

Another widely used effective Lagrangian
is the Chiral Lagrangian.  It assumes
a heavy Higgs boson using a non-linear realization of the
Goldstone bosons and assumes a custodial $SU(2)$
\cite{bagger93,appelquist80,longhitano81}.
\begin{eqnarray}
{\cal L}_1 = {L_1\over 16\pi^2} [Tr(D^\mu \Sigma^\dagger D_\mu
\Sigma)]^2 \\
{\cal L}_2 = {L_2\over 16\pi^2} [Tr(D^\mu \Sigma^\dagger D_\nu
\Sigma)]^2
\end{eqnarray}
where $\Sigma=\exp(iw^i\tau^i/v)$,  $v=246$~GeV, and
$D_\mu \Sigma = \partial_\mu \Sigma + {1\over 2} i g W_\mu^i \tau^i \Sigma
-{1\over 2} i g' B_\mu \Sigma \tau^3 $.
In this approach ${\cal L}_{1,2}$ would be the most important
manifestation of
alternative symmetry breaking scenarios in a Higgsless world.

The two approaches are not distinct so that
${\cal L}_{1,2}$ is equivalent to ${\cal L}^{o,c}_4$ with the
mapping:
\begin{equation}
g_{o,c}= {e^2 \over 16\pi^2} {1\over s_w^2} L_{1,2}
\end{equation}

Typical models with Goldstone bosons interacting with a scalar,
isoscalar resonance like the Higgs boson give $L_i \sim {\cal O}(1)$
\cite{dawson93}.
{}From precision measurements of the $Z^0$ widths Dawson and Valencia
obtained the weak bounds $-28 \leq L_1 + \frac{3}{2} L_2 \leq 26$
\cite{dawson95}.  Imposing perturbative unitarity gives the rough
constraints of $|L_1|\leq 0.3$ \cite{cheung94b}.
Therefore, the genuine quartic couplings are presently not well
constrained by experiment but are limited by perturbative
unitarity.  To facilate comparison of different processes I have
presented results in terms of $L_{1,2}$, rescaling results where
necessary (using $\alpha=1/128$ and $\sin^2_w=0.23$).  I have
defined $L_s$ when $L_1=-L_2$.

Similarly, one can write down operators involving two photons
in the Chiral Lagrangian\cite{baillargeon}:
\begin{eqnarray}
{\cal L}_o^{2\gamma} & = & -{L_o^{2\gamma}\over \Lambda^2} \left\{
K_o^W g^2 Tr(W_{\mu\nu}W^{\mu\nu}) + K_o^B g'^2
Tr(B_{\mu\nu}B^{\mu\nu}) \right. \\
& & \left. \qquad +K_o^{WB}gg'
Tr(W_{\mu\nu}B^{\mu\nu})\right\} Tr(D^\alpha \Sigma^\dagger D_\alpha
\Sigma ) \\
{\cal L}_c^{2\gamma} & = & -{L_c^{2\gamma}\over \Lambda^2} \left\{
K_c^W g^2 Tr(W_{\mu\alpha}W^{\mu\beta}) + K_c^B g'^2
Tr(B_{\mu\alpha}B^{\mu\beta}) \right. \\
& & \left. \qquad  +K_o^{WB}gg'
Tr(W_{\mu\alpha}B^{\mu\beta})\right\} Tr(D^\alpha \Sigma^\dagger
D_\beta \Sigma )
\end{eqnarray}
where
$W_{\mu\nu}  =  {\tau^i\over 2} (\partial_\mu W_\nu^i - \partial_\nu
W_\mu^i -g \epsilon^{ijk} W_\mu^i W_\nu^k ) $
and
$B_{\mu\nu}  = {1\over 2} (\partial_\mu B_\nu - \partial_\nu
B_\mu)\tau_3 $.

For $\gamma\gamma$ reactions, by making explicit the $U(1)_{QED}$
symmetry, gives the mapping:
\begin{equation}
a_{o,c}={4e^2\over s_w^2} L_{o,c}^{2\gamma} (K_{o,c}^W +K_{o,c}^B
+K_{o,c}^{WB})
\end{equation}

\section{Measurement of Quartic Couplings}

In this section I survey the various processes that have been
proposed to measure quartic couplings.

\subsection{Measurement of the Dimension 4 Operators}

\subsubsection{The Processes $e^+e^-\to W^+W^-Z, \; ZZZ$}

%\subsubsection{Triple Gauge Boson Production in $e^+e^-$}

At an 500~GeV $e^+e^-$ collider
the $W$ fusion process will be ineffective so that
three gauge boson production may be  a reasonable
substitute for the measurement of quartic couplings.
In the process $e^+e^- \to VVV$ four $W$ quartic couplings don't
contribute so that vertices
with at least two neutral vector bosons where one of
the neutrals couples to the $e^+e^-$ vertex
are likely to be the best tested in $e^+e^-$ collisions.
For any model with $SU(2)$,
however, $WWWW$ vertices are related to $WWZZ$.  $SU(2)$ also
predicts a $4Z$ vertex which will contribute to $e^+e^-\to ZZZ$.
The event rates for
reactions that meet this criteria are shown in Table \ref{table1}
for $\sqrt{s}=500$~GeV and assuming an integrated luminosity of
L=10~fb$^{-1}$.  The approach used is to look for deviations in the
cross sections from their standard model values
\cite{belanger92a,grosse93}.

\begin{table}
\caption[]{ Event rates for various $VVV$ final states in the reaction
$e^+e^- \to VVV$ for $\sqrt{s}=500$~GeV and L=10~fb$^{-1}$. From
B\'elanger and Boudjema Ref. \cite{belanger92a}.}
\label{table1}
\begin{tabular}{lrl}
Final State & Events & Comments \\
\tableline
$WWZ$ 	& 400 & $M_H < 2M_W$ or $M_H >1$~TeV \\
	& 460 & $M_H=200$~GeV \\
$ZZZ$	& 9 & $M_H>1$~TeV \\
%$WW\gamma$ & 1356 & $\theta_{\gamma beam}> 15^o$ \\
%$ZZ\gamma$ & 147 & $p_{T\gamma}>20$~GeV \\
%$Z\gamma\gamma$ & 465 & $2\gamma$'s separated by 15$^o$ \\
$ \begin{array}{l} WW\gamma \\ ZZ\gamma \\ Z\gamma\gamma \end{array} $ &
$ \begin{array}{r} 1356 \\ 147 \\ 465 \end{array} $ &
$ \left\{ \begin{array}{l}
\theta_{\gamma beam}> 15^o  \\
p_{T\gamma}>20 \mbox{ GeV} \\
2\gamma \mbox{'s separated by 15}^o \end{array} \right. $ \\
\end{tabular}
\end{table}

The process $e^+e^- \to W^+ W^- Z$
involves TGV's, QC's and Higgs exchange. In the
standard model there is a subtle cancellation between the various
contributions.   Only
anomalous QC's were considered under the assumption that TGV's can be
measured better elsewhere and assume the large $M_H$ limit so that Higgs
exchange can be neglected.    The standard model cross section
is 39.88~fb.  In their analysis B\'elanger and Boudjema
included the 67\% BR corresponding to the $6\; jet$ and
$4\; jet + e^\pm$ or $\mu^\pm$ final states, not including $\tau$'s.
The signal can be enhanced by using right
handed electrons.  The cross section is shown in Fig. 1 as a
function of $L_1$.  The most dramatic effects are for
longitudinal $W$'s with virtually no sensitivity in the TTU mode.
The limits obtained on the couplings are based on a $3\sigma$
deviation in the total unpolarized cross-section including the 67\%
BR defined above and only taking statistical errors into account.
One obtains the sensitivities\cite{belanger92a}:
\begin{eqnarray}
-96 &  < L_1 < & 81 \nonumber \\
-120 & < L_2 <  & 120 \nonumber \\
-81 & < L_s < & 70
\end{eqnarray}
One could use distributions to distinguish between $g_o$ and $g_c$.
It turns out the the $E_Z$ distribution is especially good at this.

\begin{figure}%[ht] % fig 1
%\vspace*{2.25in}
\centerline{\psfig{file=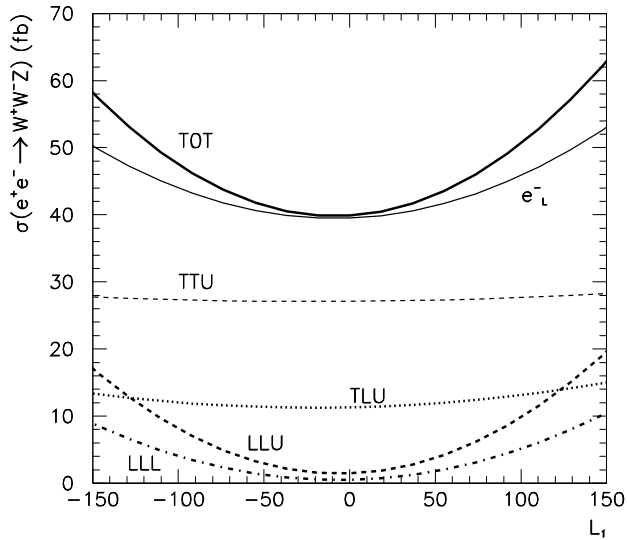,height=7.5cm}}
\caption[]{Cross-section for $e^+e^- \to W^+W^-Z$ as a function of
$L_1$ for $\sqrt{s}=500$~GeV.
Shown are the total unpolarized (TOT) cross-section, with left-handed
electrons $e_L^-$ and unpolarized cross-sections
for various combinations of vector boson polarizations: T
for transverse, L for longitudinal and U for unpolarized.
The third label is for the $Z$ polarization. From B\'elanger and
Boudjema, Ref. \cite{belanger92a}. }
\end{figure}

For the process $e^+e^- \to ZZZ$
the only SM contribution is via the Higgs boson so
that the SM cross section is very small, $\simeq 1$~fb,
making it very sensitive to anomalous couplings.  Here B\'elanger and
Boudjema consider $6 \; jet$ and $ 4 jet + \not{E}$ (not including
$\tau$ final states) corresponding to 87\% of events\cite{belanger92a}.
To use these modes one will need good invariant
mass reconstruction to distinguish from the $WWZ$ final states.  The
largest deviations are seen in the LLU channels.
Because the event rate is so small they impose the need for 50 $ZZZ$
events which gives the bounds:
\begin{equation}
-78 < L_1, \; L_2 < 85.
\end{equation}
Using a less conservative, naive,
$4\sigma$ deviation from the SM corresponding to 12 events gives
\begin{equation}
-44 < L_1, \; L_2 < 48.
\end{equation}
If deviations were observed, comparing the deviations in the $ZZZ$
mode to those found in the $WWZ$ mode could be used to find the
nature of the QC's.

\subsubsection{The Processes $e^-e^-\to VV' ff'$}

$e^-e^- \to VV' ff'$ is another option that has been examined
\cite{cuypers94,bilchak88,kuroda88}.  It
has the advantage of no hadronic background and a low SM cross
section due to the cancellation of diagrams.  The reactions
considered are:
\begin{eqnarray}
e^-e^- 	& \to & e^-e^- Z^0 Z^0 \nonumber\\
	& \to & e^-\nu_e Z^0 W^-  \nonumber\\
	& \to & \nu_e \nu_e W^- W^-
\end{eqnarray}
Note that in the last reaction only the combination $g_o+g_c$
($L_1+L_2$) can be probed.
Cuypers and Kolodziej \cite{cuypers94} performed an analysis assuming
an integrated luminosity of
10~fb$^{-1}$ and included a 1\% systematic error.  They used a
$10^o$ cut on the primary electrons and included reconstruction
efficiences.  They obtained the 95\% C.L.
contours shown if Fig. 2.
\begin{figure}%[ht] % fig
%\vspace*{2.25in}
\centerline{\psfig{file=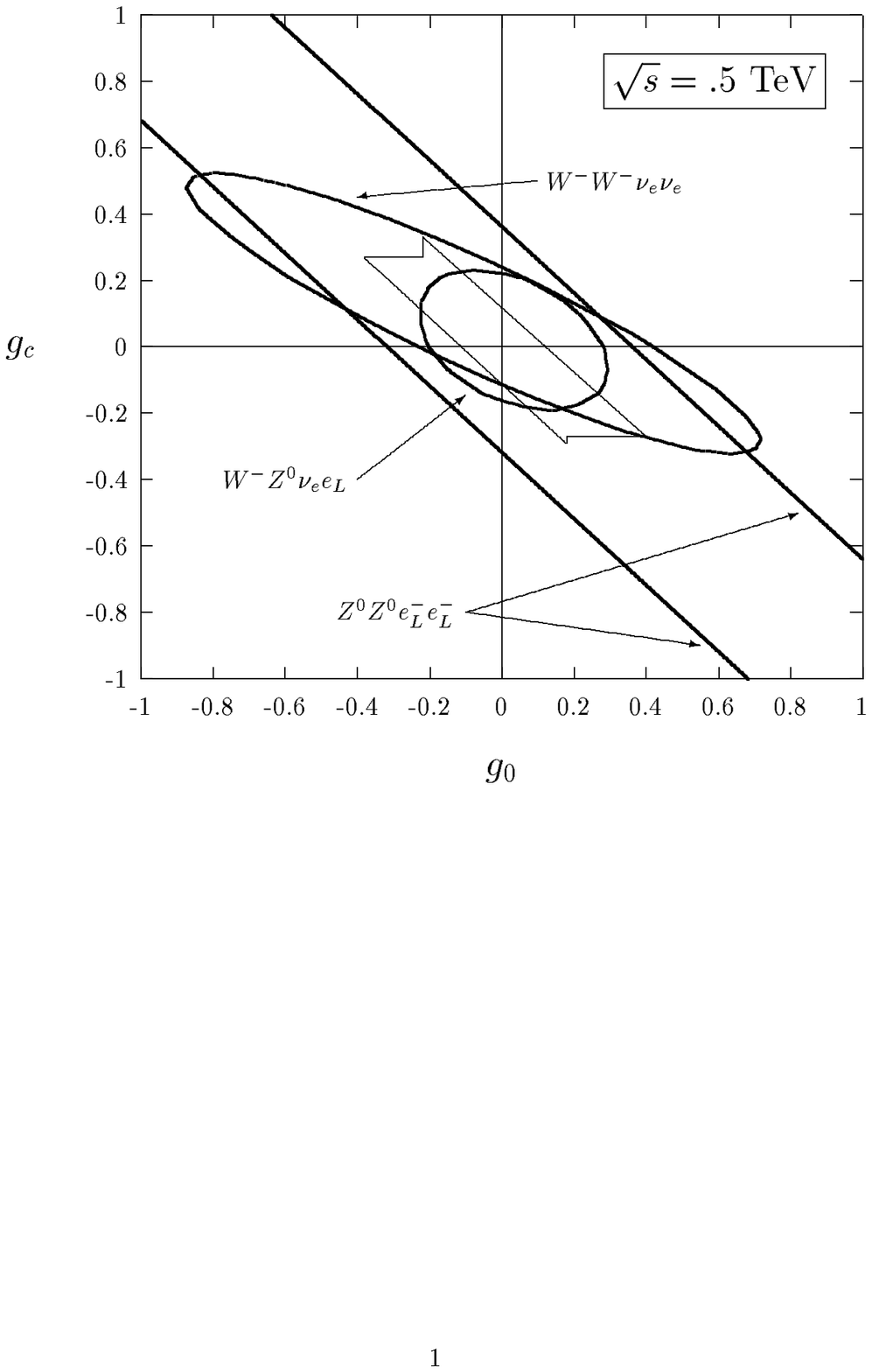,height=7.5cm}}
\caption[]{Contours of observability at 95\% C.L. of anomalous QC's
$g_o$ and $g_c$.  The measurements are for $\sqrt{s}=500$~GeV and
L=10~fb$^{-1}$.  The limits which can be obtained under similar
conditions in the $e^+e^-$ mode of the same collider are indicated
by the thin line. From Cuypers and Kolodziej Ref. \cite{cuypers94}. }
\end{figure}

\subsubsection{$W$ Fusion in $pp$ Collisions}

Although $WW$ scattering\cite{bagger93,dawson93,han93}
is covered by other contributions to these
proceedings\cite{chivukula},
it is a sufficiently important topic that a few brief comments are
included for completeness.  If the Goldstone
bosons are
non-linearly realized then one would expect new strong interactions at
$\sim$~1~TeV responsible for EWSB.  This might manifest itself as:
\begin{itemize}
\item Longitudinal $W$ states in, for example, technicolour.
\item Strong $WW$ interactions in, for example, composite scalars.
\end{itemize}

Although $W_L W_L$ can be studied in
both $e^+e^-$ and $pp$ colliders, because adequate $W_L$ luminosity
requires the highest energy possible it is best studied at the
higher energy hadron colliders.
Isoscalar resonances could be studied in $W^+ W^-$ and $ZZ$
scattering, isovector resonances in $WZ$ scattering and non-resonant
effects in $W^+W^+$.
The best channel to look for the
effect of genuine quartic couplings is the like-sign $W$ pair
production, $W^\pm W^\pm$.
Bagger {\it et al.} \cite{bagger93}
find that $pp\to W_L^+ W_L^+$ scattering at
the LHC would be sensitive to $|L_1, \; L_2|>1$.
%Bagger {\it et al.} \cite{bagger93}
%estimate that measurements at the LHC will be
%sensitive to $L_1^r(\mu_)\leq -0.75$ with the other $L_i$ set to zero
%and $\mu=1.5$~TeV.  Taking $L_1^r(\mu)=-L_2^r(\mu)$ they obtain the
%limits $-4.0\leq L_1^r(\mu) \leq -1.0$ and $L_1^r(\mu)\geq 0.8$.

\subsubsection{ $W_L W_L$ Rescattering in $e^+e^-$ and $\gamma\gamma$}

In $e^+e^-\to W W$ quartic couplings are studied
via the effects of final state interactions \cite{barklow91,han93}.
The rescattering can take place via scalar
($[I,J]=[0,0]$) Higgs like) or
vector ($[I,J]=[1,1]$ $\rho$ like) exchange.
The $W_L$'s can be related to $\pi$'s via low
energy theorems and chiral perturbation theory.  Resonance effects for
a $\rho$ like resonance are
noticible at a $\sqrt{s}=500$~GeV collider up to 5~TeV
\cite{barklow91}.
Resonances in the $I=2$ channel could be studied in $e^-e^-$.

It may also be possible to study $WW$  rescattering
at TeV energies in $\gamma\gamma$ colliders\cite{cheung94,berger94}.
Berger and Chanowitz\cite{berger94} have examined rescattering effects in
$\gamma\gamma\to ZZ$ in analogy to $\gamma\gamma\to \pi^0\pi^0$.
They concluded that the background overwhelms the
signal unless there are strong resonance effects from, for example,
an $f_{2TC}$ with mass $\sim 3.4$~TeV ($N_{TC}=3$).
A very high energy collider of $\sqrt{s_{\gamma\gamma}}=3.2$~TeV
($\sqrt{s_{e^+e^-}}=4$~TeV) with
high luminosity, of order 100~fb$^{-1}$, would be
needed to see its effects.

\subsection{Measurement of Dimension 6 operators}

\subsubsection{The Processes
$e^+e^-\to W^+W^-\gamma, \; ZZ\gamma, \; Z\gamma\gamma$}

%\paragraph{$e^+e^- \to W^+ W^- \gamma$\cite{grosse}}

The process $e^+e^- \to W^+ W^- \gamma$
is used to study the $W^+W^-\gamma\gamma$ and $W^+W^-
Z\gamma$ couplings.  It has the largest cross section of all 3-boson
production and is quite sensitive to dimension 6 operators.  The
largest deviations occur when both $W$'s are longitudinal.  B\'elanger
and Boudjema \cite{belanger92a} impose the cuts $P_{T\gamma}>20$~GeV,
$\theta_{e\gamma}>15^o$, and $|\eta_\gamma|< 2$ resulting in a cross
section of $\sigma_{WW\gamma}=135.6$~fb.  They used the 79\% of the
BR that does not include $\tau$'s
with 45\% being $4\; jet + \gamma$. The anomalous QC's contribute
significantly to cross-sections with right handed electrons.
They obtain the $3\sigma$ limits:
\begin{eqnarray}
-62 &  < a_o < & 93 \nonumber \\
-110 & < a_c < & 47
\end{eqnarray}
The different operators give different distributions for $E_\gamma$
but not for $\theta_{\gamma W}$.
$e^+e^- \to ZZ\gamma$ has a SM cross-section of 14.7~fb with the
same cuts as above.
It turns out that constraints obtained
from $e^+e^- \to ZZ\gamma$ are less constraining than the
$WW\gamma$ final state and
%that anomalous quartic couplings
%as large as 1 have virtually no effect on the
the $e^+e^- \to Z\gamma\gamma$ cross section is very insensitive to
anomalous couplings.

Leil and Stirling \cite{leil94} used the process $e^+e^-\to
W^+W^-\gamma$ to study $a_n$.  They imposed the cuts
$|\eta_\gamma| \leq 2$, $E_\gamma >20\%$ to avoid collinear
singularities and particle separation of
$15^o$, obtaining a cross section for $\sqrt{s}=500$~GeV
of $\sigma_{SM}=123.4$~fb.  Using
the $E_\gamma$ spectrum they obtain the additional limit
based on L=10~fb$^{-1}$ and requiring $3\sigma$ deviations
% , and assuming $\Lambda=M_W$
of
\begin{equation}
-610 < a_n < 660.
\end{equation}

\subsubsection{The Processes
$\gamma\gamma \to W^+W^-$ and $\gamma\gamma\to ZZ$ }

These reactions are in the pure non-abelian gauge sector of the SM.
Both the trilinear and quartic couplings enter.  Since the TGV's can
be constrained better elsewhere these reactions are ideal tests of
the quartic couplings.
The $WW\gamma\gamma$ and $ZZ\gamma\gamma$ couplings are related by
$SU(2)$ but because they contribute to different observables we can
set independent bounds on them.

The process $\gamma\gamma \to W^+W^-$ constitutes the largest
cross-section in $\gamma\gamma$ collisions, with a cross-section
at 400~GeV of
$\sigma=80$~pb making a $\gamma\gamma$ collider a $W$-Factory.  The
angular distributions are shown in Fig. 3.
The SM contributions are peaked along the initial photon directions
while the anomalous QC's are more central.
Even with angular cuts the SM contributions are still large.  The photon
helicities can be used to separate different contributions.
In the $\lambda_1=\lambda_2$ mode ($J=0$) the SM does not produce
$W$'s of different helicities.   This is maintained for $a_o$ so that
$a_o$ only contributes to $J=0$ while $a_c$ contributes to both $J=0$ and
$J=2$.  Because $a_o$ and $a_c$ have the same S-wave amplitudes
distinguishing them requires the use of the photon helicity
amplitudes.
Taking $\cos\theta<0.7$ the SM cross-section is 17.58~pb so that
statistical errors are negligible and the main source of error is
systematics.  For L=10~fb$^{-1}$ and assuming $\Delta
\sigma /\sigma = 3\%$ B\'elanger and Boudjema \cite{belanger92b} obtain:
\begin{eqnarray}
%-5\times 10^{-2} &  < a_o < & 2\times 10^{-2} \quad J_Z=0 \nonumber \\
%-0.1 & < a_c < & 3.6 \times 10^{-3} \quad J_Z=0 \nonumber \\
%-2\times 10^{-2} & < a_c <  & 2\times 10^{-2} \quad J_Z=2
-7.8 &  < a_o < & 3.1 \quad J_Z=0 \nonumber \\
-16 & < a_c < & 0.56 \quad J_Z=0 \nonumber \\
-3.1 & < a_c <  & 3.1 \quad J_Z=2
\end{eqnarray}
Ratios could also have been used which eliminates the need to
measure the $\gamma\gamma$ luminosity.  Using angular
distributions could give additional information.

\begin{figure}%[ht] % fig 2
%\vspace*{8.0cm}
%\begin{center}
%\vskip -8.0cm
%\begin{turn}{90}
%\mbox{\epsfig{figure=ggww.ps,height=8.0cm}}
%\end{turn}
%\end{center}
\centerline{\psfig{file=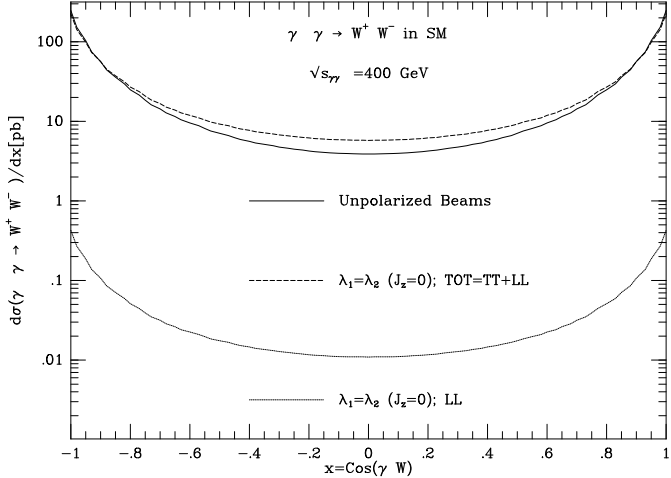,height=8.0cm}}
\caption[]{$W$ angular distribution in the process $\gamma\gamma\to
W^+W^-$ at $\sqrt{s}=400$~GeV for different initial photon
helicities. From B\'elanger and Boudjema, Ref. \cite{belanger92b}.}
\end{figure}

%\subsection{$\gamma\gamma \to ZZ$}

The process $\gamma\gamma \to ZZ$ is attractive as
the SM background is very small.
$SU(2)$ relates
the $ZZ\gamma\gamma$ vertex to the $WW\gamma\gamma$ vertex so
combining this and the previous reaction is an ideal way of
testing for $SU(2)$ symmetric QC's.  As before,
$a_o$ contributes to $J_z=0$
while $a_c$ contributes to both the $J_z=2$ and $J_z=0$ channels
making it possible to distinguish the 2 quartic couplings.  The
$J_Z=0$ and $J_Z=2$ channels can be distinguished using polarization
and angular distribution information.

Unfortunately, in the original analysis of these process it was assumed that
the SM
cross section was zero.  A subsequent 1-loop calculation by Jikia
\cite{jikia} found that, although small, the SM cross section was not
neglible and was dominated by the transverse modes. Nevertheless it
is believed that properly including the SM contribution will still
result in useful bounds, in much the same way as the
$\gamma\gamma\to W^+W^-$ case \cite{baillargeon}. In the absence of
a detailed analysis we describe the estimate of Baillargeon
{\it et al.} \cite{baillargeon}.  The SM $Z_L Z_L$ contribution in
the heavy Higgs mass limit  is quite small at all energies,
$\sim 1$~fb.  Therefore to obtain a crude estimate as to how the
limits are changed it is sufficient to include the SM $Z_T Z_T$
contribution that is not affected by anomalous QC's.  The limits
are based on the total cross section only. One could exploit
the fact that the $TT$ cross section is relatively insensitive to
the $J_Z$ of the initial two photons to construct an asymmetry such
as $\sigma(J_Z=0) - \sigma(J_Z=2)$ to reduce the SM background.
This, of course assumes that the new physics does not contribute
equally to the two $J_Z$.  Baillargeon {\it et al.} include only the
visible, unambiguous $ZZ$ signal with one $Z$ decaying hadronically
and the other leptonically with the cut $\cos\theta_Z < 0.866$. The
criteria of observability was based on requiring $3\sigma$
statistical deviation from the SM cross- section.
% Taking $\Lambda=1$~TeV they obtain:

%\begin{array}{llll}
%|a_o|<2 & |a_c|< 5 & \sqrt{s_{ee}}=500\; \mbox{GeV} & L=10\; \mbox{fb}^{-1} \\
%|a_o|<0.3 & |a_c|< 0.7 & \sqrt{s_{ee}}=1\; \mbox{TeV} & L=60\; \mbox{fb}^{-1}
%\end{array}

%\begin{tabular}{llll}
%$|a_o|<2$ & $|a_c|< 5$ & $\sqrt{s_{ee}}=500$ GeV & L=10 fb$^{-1}$ \\
%$|a_o|<0.3$ & $|a_c|< 0.7$ & $\sqrt{s_{ee}}=1$ TeV & L=60 fb$^{-1}$
%\end{tabular}

\begin{eqnarray}
|a_o|<2 \qquad & |a_c|< 5 \qquad & \qquad
	(\sqrt{s_{ee}}=500 \mbox{ GeV  L=10 fb}^{-1})\nonumber \\
|a_o|<0.3  \quad & |a_c|< 0.7 \quad & \qquad
	(\sqrt{s_{ee}}=1 \mbox{ TeV  L=60 fb}^{-1} )
\end{eqnarray}

%\begin{figure}%[ht] % fig 3
%\vspace*{2.25in}
%\centerline{\psfig{file=fig1.eps,width=12.0cm}}
%\caption[]{$W$ angular distribution in the process $\gamma\gamma\to
%ZZ$ at $\sqrt{s}=400$~GeV for different initial photon
%helicities.}
%\end{figure}

\subsubsection{The Processes $\gamma\gamma \to W^+W^- Z$
and $\gamma\gamma \to W^+W^- \gamma$}

\'Eboli {\it et al.}, \cite{eboli95} have studied the the processes
$\gamma\gamma \to W^+W^- Z$ and $\gamma\gamma \to W^+W^- \gamma$.
They found that the constraints from the first reaction on $a_n$ is
as restrictive as the contraint obtained in $e\gamma$ collisions.
The limits on $a_c$ are an order of magnitude better than those
coming from the $e^+e^-$ mode  and are comparable to the limits that
can be obtained in the $e\gamma$ mode.  However, they are a factor
of 2 weaker that those obtained from $\gamma\gamma \to W^+ W^-$.
The limits on $a_o$ are slightly better than those obtained in the
$e^+e^-$ mode but an order of magnitude worse than those obtained in
$e\gamma$ or $\gamma\gamma \to W^+W^-$.

\subsubsection{The Processes $e\gamma \to V V' f$}

A number of authors have studied the effects of anomalous QC's on
the reactions $e\gamma\to VV' f$ \cite{eboli94,cheung94b,rosenfeld94}.
The cross sections are summarized
in Fig. 4 which gives the cross sections in $e\gamma$ collisions as
a function of $\sqrt{s}$ \cite{cheung93}.
The $WWe$ and $ZZe$ final states are most
sensitive to $a_o$ and $a_c$ although $WWe$ is insensitive to $a_n$.
The cross section $\sigma(WWe)$ is an
order of magnitude larger than $\sigma(WW\gamma)$ due to t-channel
photon exchange.  Likewise, for $ZZe$ t-channel photon exchange is
introduced by the $ZZ\gamma\gamma$ vertex, not present in the SM,
making it a very sensitive
process.  The results from an analysis of \'Eboli {\it et al.}
\cite{eboli94}
for the sensitivies of $e\gamma \to VV'f$ to the various QC's,
are summarized in Table 2.  Their results are based on $3\sigma$
effects based on statistics for 10~fb$^{-1}$ integrated luminosity.
The conclusion of these studies is that
they are not quite as good as those coming from
$\gamma\gamma$ reactions for the study of QC's except for $a_n$
which requires the smaller phase space process
$\gamma\gamma\to W^+ W^- Z^0$ in $\gamma\gamma$ collisions.

\begin{figure}%[ht] % fig 3
%\vspace*{2.25in}
\centerline{\psfig{file=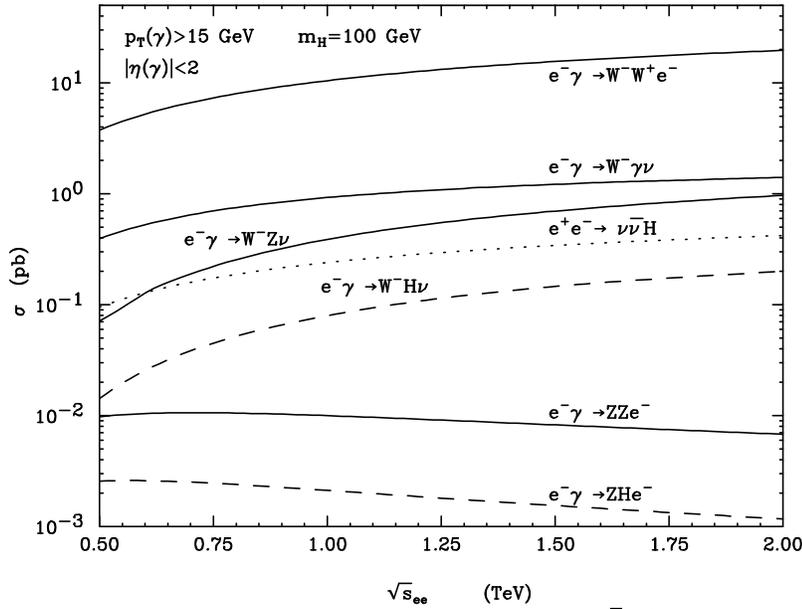,height=8.0cm}}
\caption[]{Cross sections for $e\gamma$ processes
as a function of $\sqrt{s}$ with the acceptance
cuts $p_T(\gamma) > 15$~GeV and $|\eta(\gamma)|<2$.
{}From K. Cheung  Ref. \cite{cheung93}.  }
\end{figure}

\begin{table}
\caption[]{Sensitivities of $a_o$, $a_c$ and $a_n$ to $e\gamma \to VV' f$
corresponding to $3\sigma$ deviations, varying one coupling at a time.
For events containing a
photon in the final state the cut $p_{T\gamma}>15$~GeV was used to
eliminate collinear divergences. From \'Eboli {\it et al.} Ref.
\cite{eboli94}.  }
\label{table2}
\begin{tabular}{lccc}
Final State & $a_o$ & $a_c$ & $a_n$ \\
\tableline
%$WWe$ 	& $-0.21<a_o<0.036$ & $-1.5<a_c<1.4$ & $-4.5<a_n<4.5$ \\
%$Z\gamma e$ 	& $-0.94<a_o<0.95$ & $-1.3<a_c<1.4$ & --- \\
%$ZZe$ 	& $-0.029<a_o<0.028$ & $-0.098<a_c<0.095$ & --- \\
%$W\gamma\nu$ & $-0.56<a_o<0.54$ & $-0.57<a_c<1.1$ & --- \\
%$WZ\nu$	& ---	& --- & $-1.2<a_n<0.74$ \\
$WWe$ 	& $-33<a_o< 5.6$ & $-230<a_c<220$ & $-700<a_n<700$ \\
$Z\gamma e$ 	& $-150<a_o<150$ & $-200<a_c<220$ & --- \\
$ZZe$ 	& $-4.5<a_o<4.4$ & $-15<a_c<15$ & --- \\
$W\gamma\nu$ & $-87<a_o<84$ & $-89<a_c<170$ & --- \\
$WZ\nu$	& ---	& --- & $-190<a_n<120$ \\
\end{tabular}
\end{table}

\section{Summary}

One of the most important problems in particle physics is the
understanding of electroweak symmetry breaking.
If the Higgs boson is ``heavy'' and electroweak symmetry breaking is
non-linearly realized then the quartic vertices will provide a window
into EWSB.  I have surveyed various processes that have been
proposed to study quartic couplings.  For dimension 4 operators
$e^+ e^- \to W^+W^- Z, \; \to ZZZ$, $e^-e^- \to VV' ff'$, $pp\to W W
+X$, and $WW$ rescattering in $e^+e^-\to W^+W^-$ have been
considered and for dimension 6 operators $e^+e^- \to W^+W^-\gamma, \;
ZZ\gamma$, $\gamma\gamma\to W^+W^- ,\; ZZ$, and $e\gamma \to
WW\gamma ,\; ZZ\gamma , \; WZ\nu$.

For the dimension 4 operators it appears that the LHC will
provide the most constraining measurements.  However,
is far from clear whether the LHC will be able to
disentangle this sector of the weak interaction.  It is
possible, then, that precision measurements at high energy $e^+e^-$
colliders through $W^+W^-$ rescattering
could be our first glimpse of a strongly interacting weak interaction.

For the dimension 6 quartic couplings involving
photons, $\gamma\gamma$ collisions appear to be the best place to
measure these couplings.  They can be measured at least an order of
magnitude more precisely than using 3-boson production in $e^+e^-$.
Measurements using gauge boson production in $e\gamma$ collisions are
almost as precise as the $\gamma\gamma$  processes.

The study of quartic couplings are still in the preliminary stages.
It would be useful for the purposes of comparing
different processes that a consistent parametrization of the
vertices be adopted and that the different processes be analysed in
a consistent way.
The most dramatic effect of QC's is when all vector bosons are
longitudinal.  Therefore, an important next step is to
include the decays of the $W$'s and $Z$'s into fermions
and their reconstruction.   After all, it is the fermions which are
observed not the gauge bosons themselves.  This would enable more
sophisticated polarization studies that would simulate
the experimental separation
of $W$'s and $Z$'s and the separation of longitudinal and transverse
gauge bosons.

\acknowledgments

The author is most grateful to
Genevieve B\'elanger for many helpful communications in
preparing this review and a careful reading of the manuscript.
The author thanks Genevieve B\'elanger, Fawzi Boudjema,
Kingman Cheung, and Frank Cuypers for graciously supplying him with the
figures included here.
The author thanks the organizers of TGV95 for their kind
invitation to attend a most
enjoyable meeting and the Deans of Research and Science at Carleton
University  for financial support to attend the meeting.
This research was supported in part by the Natural Sciences and
Engineering Research Council of Canada.

\end{document}